\documentclass{article}
\begin{document}
\title{Varying Newton constant,entropy and the black hole evaporation law}

\author{Julia Haba\thanks{email:haba.jula@gmail.com}
\and Zbigniew Haba
\thanks{email:zbigniew.haba@uwr.edu.pl}}
\date{Institute of Theoretical Physics , University of Wroclaw\\50-204 Wroclaw, Plac Maxa Borna 9, Poland}
\maketitle

\begin{abstract}
We assume that the Newton potential of a spherically symmetric body of mass  $M$ is 
$V=-G(M)Mr^{-1}$, where  the Newton constant $G$ may depend on
 $M$. We insert this assumption in the Schwarzschild solution. 
  We discuss  varying $G$ in the black hole thermodynamics and black hole evaporation law.
We extend the Bekenstein-Hawking formula  $\hbar GS=\frac{1}{4}A$, where $S$ is the entropy and $ A$ the area, to a varying $G$. It is shown that if $G(M)$ is of the form of a power-law $G\simeq M^{-\Gamma}$ then $G(M)S\simeq A$ is still preserved but
   the evaporation law substantially depends on the function $G(M)$.

\end{abstract}
 \maketitle
\section{Introduction}
A measurement of the Newton constant $G$ is achieved with still unsatisfactory precision of 
$G^{-1}\delta G\simeq 10^{-5}$ \cite{g}\cite{g1} ( even less is known  for very small and large distances).
An eventual variation of $G$ with respect to  some parameters of the system is even harder to determine through astronomical observations. $G$ enters astrophysical results usually in a product with the mass $M$. Then, the variation of $GM$ can be established with much higher precision \cite{uzan} (usually with an assumption of a constant $M$).
The Schwarzschild solution of Einstein equations is determined up to a constant which is fixed as
$GM$ by the requirement that the gravitational potential (at a weak gravitational field experienced at large distances from a spherically symmetric compact  body) should coincide with the one of Newton theory.
In this paper we admit that $G$ in the Newton potential may depend on $M$. We study some consequences of such a dependence for thermodynamics of the black hole ( with no angular momentum and no charge)
and its evaporation law. There is an indication for a possible $M$-dependence of $G$ in some theories.  A dependence of $G$ on a scalar field $\phi$ interacting with gravity appears in scalar
tensor theories of gravity . The first model of this type is the one of Brans-Dicke \cite{dicke}
( for an application to black holes see \cite{barrowcarr}\cite{carr}). The dependence of $G$ on $M$
in such models
could appear as a rough approximation. In models with $G$ and $M$ depending on time \cite{gupta}\cite{costa}\cite{blanchard} we can express the time in $G$ by $M$.
We discuss thermodynamics of the black hole with the Newton constant depending on $M$.
As a prospect for a preservation of the Bekenstein-Hawking  formula for the black hole entropy $S$ we consider the supposition that the energy $E$ which enters the first law of thermodynamics
$dE=TdS$ (where $T$ is the temperature) is equal  to $q M$ with the Hawking mass $M$ and a certain $q\neq 1$. The black hole area $A$ and temperature depend on $GM$. Then, the Bekenstein-Hawking entropy multiplied by $G$ is proportional to the area.
In sec.2 we show that if $G\simeq M^{-\Gamma}$
then $GS\simeq 16\pi G^{2}M^{2}$ is still proportional the black hole area.
If $\Gamma=2(1-q)$ (with a constant $q$) then the Bekenstein-Hawking formula for entropy  remains unchaged (but the black hole area depends on $G(M)$).
We discuss briefly an eventual
dependence of $q$ on $M$.
In sec.3 we investigate a loss of energy by an evaporation of massless particles. The evaporation law depends on the area and the
temperature of the black hole. We assume the standard Hawking formula for the temperature and the area
 as a function of $GM$. We calculate the decrease of the mass as a function of time.
The dependence of $G$ on $M$ leads to some  differences with respect to the result of Page and Hawking \cite{pagehawking}.
The black hole life time is longer and the mean density of the black hole can remain finite ( that a
varying $G$ could change the evaporation law has been discussed in
\cite{barrowcarr}\cite{carr}).
 The evaporation law predicts that if $\Gamma<1$ then  the temperature and luminosity
 tend to infinity at the end of the evaporation. This can change if $\Gamma\geq 1$ or if $q$ is $M$-dependent.
 However, such systems seem rather exotic.

 The compact systems with an event horizon could lead to
  useful tests (besides the ones discussed in \cite{uzan})
  of a varying Newton constant $G$ because the area and the
  temperature of the horizon depend on $G$. The knowledge of an exact
evaporation law may be important for recent   searches of the
primordial black hole explosions \cite{evapobs2}\cite{evapobs3}
raised by an observation of a giant emission of neutrino radiation
in KM3NeT \cite{evapobs1} (for a search  of radiation from
primordial black holes see \cite{burst} and references cited
there).

\section{Thermodynamics with a varying $G$}

Under the assumption of a spherical symmetry the static asymptotically flat vacuum 
solution of  Einstein equations outside a compact body is $g_{00}=-(g_{11})^{-1}=1-Kr^{-1}$, where $K$ is a constant. The agreement with the Newton law at large distance from the body of mass $M$ determines $K=2GM$
(we choose units with the velocity of light $c=1$) because in a linearized Einstein gravity $\frac{1}{2}(g_{00}-1)$ is the Newton potential.
This leaves open the option that $G$ may depend on $M$.  In most applications of gravity to astronomy
$G$ enters in the product $GM$. For this reason $GM$ is measured
with a much higher precision than $G$. The physics
of a static spherically symmetric black hole depends solely on 
$GM$, but nothing forbids $G$ to depend on $M$. If we find a process which depends separtely on $M$ then a study of this process will allow us to determine eventual  dependence  of $G$ on $M$.
The radius  of the spherically symmetric black  hole is
determined by the zero of $g_{00}$. Hence, the black hole area is $A=16\pi G^{2}M^{2}$.
As conjectured by Bekenstein \cite{bekenstein} the entropy of the black hole should be
 proportional to its area. 
 By dimensional reasons we must divide $A$ by $\hbar G$ in the formula for the entropy
  ($\hbar$ is the Planck constant, we choose the units with the velocity of light $c=1$ and the Boltzmann constant $k_{B}=1$). Hence, the Bekenstein conjecture is 
 
 \begin{equation}
\hbar  GS_{\kappa}=\frac{\kappa}{4} A,\end{equation} where $\kappa$ is a dimensionless constant.

There are good reasons to expect that the entropy should be proportional to the area (when $G$ is a constant).
As proved by Hawking the area of the black hole is increasing during a merger of black holes. The entropy calculated in the semiclassical approximation of the functional integral
\cite{gibbons1}\cite{gibbons2}\cite{qent2}\cite{qent3}\cite{qent1} is also proportional to the black hole's area. A microscopic counting of states
in loop quantum gravity \cite{rovelli}\cite{lewandowski}\cite{meissner} leads to the entropy (1) proportional to the area but it
does not fix the
coefficient  in front of the area . Different coefficients multiplying the area
may  arise in effective theories of gravity (some of them resulting from an interaction with quantum  matter fields \cite{barrowcarr}\cite{carr}
\cite{qent1}
\cite{davies}\cite{davies2}
\cite{faraoniBD} \cite{kang}).

 A calculation of the temperature of the black hole
in a system of quantum scalar particles on the Schwarzschild
background \cite{hawking2} gave the result (the formula has been
established for various quantum systems defined on the
Schwarzschild background \cite{wald})
\begin{equation} T_{H}=\frac{\hbar }{8\pi GM}.
\end{equation}
The formula for the temperature $T_{H}$ can also be derived directly
from the Schwarzschild metric as the surface gravity and from
the periodicity of the metric when continued to an imaginary time \cite{wald}
\cite{gibbons2} (through the KMS condition).
Then, for a constant $G$ the first law of thermodynamics  $TdS=dM$ gives the Bekenstein-Hawking formula for the entropy ( $\kappa=1$ in eq.(1))
\begin{equation}
\hbar GS_{BH}=4\pi M^{2}G^{2}=\frac{A}{4}.
\end{equation}
The first law of thermodynamics for the black hole derived in classical Einstein gravity \cite{bardean} 
identifies the energy $E$ in the formula $dE=TdS$  with $M$.
 In thermodynamics it means that  the heat supplied to the mass $M$
 will be used entirely for the increase of  this mass. In another formulation,  the  energy of the body of the Schwarzschild mass $M$ is equal to its thermodynamic energy. We think that this is not obvious
 especially in quantum systems when gravity is induced by quantum fields interaction. We suggest to express the first law of thermodynamics for the black hole in the form
\begin{equation}
dE=d(qM)=T_{H}dS,
\end{equation}
where in general $q$ and $G$ may depend on $M$  ( the physics of the chargeless spherically symmeric black hole can depend only on its mass).
An integration of eq.(4) gives
\begin{equation}S=\frac{8\pi}{\hbar} F(M),
\end{equation}
where
\begin{displaymath}
F(M)=\int G(M)M d(qM)
\end{displaymath}
In general, $GS$ in eq.(5) is not proportional to the black hole area.
It is shown in the Appendix that the requirement 
$GS\simeq A$ (as in eq.(1)) for  $q=const$ is possible only if
$G(M)\simeq M^{-\Gamma}$ with a certain real constant $\Gamma$. In such a case we obtain
\begin{equation} 
GS=\frac{2q}{2-\Gamma}GS_{BH}, 
\end{equation} if $\Gamma\neq 2$.
For $\Gamma=2$ the integral in eq.(5) gives $S\simeq \ln M\simeq -\ln A$ and $GS\simeq -A\ln A$. Hence, 
the case $\Gamma=2 $ is exceptional and should be excluded from our models
(it corresponds to the limit $\kappa\rightarrow\infty$ in the formula $\Gamma=2-\frac{2q}{\kappa}$ in eq.(17)).
If $q$ depends on $M$ then the condition $4\hbar GS=\kappa A$
(with $\kappa$ independent of $M$)
determines $G(M)$ as a function of $q(M)$ as shown in the Appendix.

An $G$ depending on $M$ when inserted into
the spherical model of a star will change the expression (the Chandrasekhar limit) for the maximal mass of the white dwarf and the neutron star.An estimate of the mass of a galaxy on the basis of the rotational velocity of stars also depends on the Newton law modified by $G(M)M$.

\section{Mass loss in evaporation}
The black hole evaporation has been discovered by Hawking \cite{hawking2} in a model
of a quantum  scalar field on a black hole background. In order to calculate the black hole mass loss through an evaporation of massless particles one can apply Planck's theory of radiation.
We restrict ourselves to the approximation that the evaporating black hole can still be described by
the $G(M)Mr^{-1}$ potential when its mass $M$ is varying in time ( see ref.\cite{bardeen} ).
The calculations have been performed by Page \cite{page2}\cite{page} (for particles of the standard model see \cite{gribbin}). It comes out (\cite{harlow}, sec.IV.D , \cite{wald2},sec.7.3, see also \cite{rio} \cite{carlip}\cite{witten}) that with a good approximation the result for the black hole energy loss coincides with the one which can be obtained from the Stefan-Boltzmann law
\begin{equation}
\frac{dE}{dt}=-\lambda T^{4}A,
\end{equation}
where  $A$ is the area of the surface. 
The coefficient ($\lambda \simeq
\hbar^{-3}$) in black hole evaporation \cite{page}\cite{gribbin} is different than in classical Stefan- 
Boltzmann model.

 We assume in all models  $T\simeq (GM)^{-1}$
and $A\simeq (GM)^{2}$. When we set $E=qM$ (in the cited papers $q=1$), insert  the temperature  and the
area in eq.(7) then
we obtain \cite{page2}
\cite{page}\cite{gribbin}\cite{pagehawking} (see also the reviews
in \cite{harlow}\cite{carlip}\cite{witten})
\begin{equation}
\frac{dE}{dt}=\frac{d(qM)}{dt}=-\hbar\alpha M^{-2}G^{-2},
\end{equation}
where $\alpha$ is a ($\hbar $ independent)  number which can be
obtained from  an estimate of
emissivity \cite{page2}. We assume
that the temperature $T$ is larger then the temperature of the
environment so that an absorption from the environment can be
neglected. Eq.(8) has been  derived under the assumption
that during the black hole evaporation the black hole's horizon is still $2GM$ and its temperature
$\hbar(8\pi GM)^{-1}$. These assumptions have been studied 
 by Bardeen \cite{bardeen} with a conclusion that they are justified if $M$ is bigger then the Planck mass. The assumptions underlying eq.(8) suggest that we should be rather cautious concernig the conclusions at a large time when the mass  is close to zero ( and the Hawking temperature of the black hole tends to infinity). Nevertheless,
 it is interesting to derive consequences of eq.(8) with
 the perspective of its possible modification ( including the option of $q$ depending on $M$).
Integrating eq.(8) we obtain
\begin{displaymath}
\int_{M_{0}}^{M}(GM)^{2}(q(M)+
\frac{dq}{dM})dM=-\alpha\hbar(t-t_{0}),\end{displaymath}
where $t_{0}$ is the time of the black hole formation. 
 
In eq.(8) we insert
\begin{equation}
G=G_{0}M_{0}^{\Gamma}M^{-\Gamma},
\end{equation}
and assume $q=const$ ($q$ dependng on $M$ is discussed briefly in the Appendix).
The Bekenstein-Hawking formula (6) ($\kappa=1$) gives $\Gamma=2(1-q) $.
We have no convincing estimates for the value of $\Gamma$ for the black hole.
Let us only point out that
if we require that the velocity $v$ of a star  in a spherically symmetric galaxy
 with a constant density $\rho$ is to be independent of its distance $r$ from the center then the 
equation for rotational velocity $v^{2}r^{-1}\simeq r^{-2}(\rho r^{3})^{1-\Gamma}$ gives
  $\Gamma=\frac{2}{3}$ .

If $G$ is a constant then integrating eq.(8) we obtain the
well-known formula
\begin{equation}
M_{t}^{3}=M_{0}^{3}-3\hbar q^{-1}\alpha G_{0}^{-2}(t-t_{0}).
\end{equation}

Let us consider the general option (9) with $\Gamma\neq 0$ . When
we substitute the relation (9)   in eq.(8)
then we obtain the solution ( when
$\Gamma\neq\frac{3}{2}$)
\begin{equation}
M=M_{0}\Big(1-(3-2\Gamma)\hbar q^{-1}\alpha
G_{0}^{-2}M_{0}^{-3}(t-t_{0})\Big)^{\frac{1}{3-2\Gamma}}\equiv
M_{0}Z^{\frac{1}{3-2\Gamma}}.
\end{equation}
Then
\begin{equation}
G=G_{0}\Big(1-(3-2\Gamma)\hbar q^{-1}\alpha
G_{0}^{-2}M_{0}^{-3}(t-t_{0})\Big)^{-\frac{\Gamma}{3-2\Gamma}}\equiv
G_{0}Z^{-\frac{\Gamma}{3-2\Gamma}}.
\end{equation}When  $\Gamma=\frac{3}{2}$ then
\begin{equation}
M=M_{0}\exp\Big(-\hbar q^{-1}\alpha G_{0}^{-2}M_{0}^{-3}(t-t_{0})\Big).
\end{equation}
In the expression for the temperature (2) and luminosity (8) ( if $\Gamma\neq \frac{3}{2}$) 
\begin{equation}\begin{array}{l}
T_{H}=\frac{\hbar}{8\pi}(GM)^{-1}=\frac{\hbar}{8\pi}(G_{0}M_{0})^{-1}Z^{-\frac{1-\Gamma}{3-2\Gamma}}
\end{array}
\end{equation}
$T_{H}$ tends to infinity if $\Gamma< 1$ and to zero if $\Gamma>1$.

The mean energy density defined as the mass divided by the volume of the black hole is 
proportional to
\begin{equation}\begin{array}{l}
G^{-3}M^{-2}=G_{0}^{-3}M_{0}^{-2}Z^{\frac{3\Gamma-2}{3-2\Gamma}}.
\end{array}\end{equation}
Note that the mean density tends to zero if $\Gamma> \frac{2}{3}$ and  is constant for $\Gamma=\frac{2}{3}$.
In
the Page-Hawking model \cite{pagehawking} the average density tends to infinity at the end of black hole's life time .

The life time $t_{f}$ of the black hole for $\Gamma<\frac{3}{2}$ according
to eq.(11) is
\begin{equation}
t_{f}=\frac{1}{3-2\Gamma}qG_{0}^{2}M_{0}^{3}\hbar^{-1}\alpha^{-1}=\frac{3q}{3-2\Gamma}t_{c},
\end{equation}where $t_{c}$ (eq.(10)) is the life time with a constant $G$ \cite{pagehawking}.
For $\Gamma\geq\frac{3}{2}$  the mass is
decreasing to zero  asymptotically in time.

The results  could be compared with observations of the
gamma-ray bursts \cite{burst}. In particular, with the recent
observations of the giant neutrino emissions
\cite{evapobs1} which by some authors
\cite{evapobs2}\cite{evapobs3} are interpreted as resulting from
primordial black hole explosions.

\section{Summary}  
Most astrophysical results are derived under the assumption that the  gravitational attraction
is universal and described by Einstein equations. However, the energy-momentum on the right side of Einstein equations depends on the system under investigation. The Schwarzschild solution outside a spherically symmetric system predicts the $r^{-1}$ gravitational potential up to a constant. The eventual departure
from this behavior requires either quantum corrections or $F(R)$ gravity. The Schwarzschild solution
admits the constant in the gravitational potential to depend on the system under consideration. The physics of a non-rotating, chargeless black hole depends only on its mass $M$. This allows the Newton constant $G$ in the  gravitational potential $G(M)Mr^{-1} $ to depend on the mass. By an observation of the motion of a test body around the black hole we would not measure $M$ but $G(M)M$. We considered some arguments for a determination of $G(M)$. If we require that the black hole entropy multiplied by $G$ is proportional to its area and $q=const$ in $E=qM$ then $G(M)$ must be of the form of a power-law $G\simeq M^{-\Gamma}$. Inserting this power-law 
behavior into the formula for the black hole evaporation we derive a result for the mass loss.
The resulting luminosity of the black hole depends on $\Gamma$ which in this way could be determined from observation.
A dependence of $q$ on $M$ in the thermodynamic energy $E$
 would futher modify the evaporation law but an eventual
form of $q(M)$ would have to be determined by a deeper insight into the quantum black hole theory. 

{\bf Appendix}

The requirement $\hbar GS=\frac{\kappa}{4}A$ with $S$ of eq.(5) defines 
an integral equation. By means of differentiation we obtain  a differential equation
\begin{equation}
\frac{d\ln G}{dm}=\frac{2}{\kappa}(q-\kappa)+\frac{2}{\kappa}\frac{dq}{dm},
\end{equation}
where $m=\ln M$. The solution is
\begin{equation}
G(M)=KM^{-2}\exp\Big(\frac{2}{\kappa}
(q(m)+\int q(m)dm)\Big),\end{equation}where $K$ is a constant.
If $q=const$ then from eq.(18) we obtain 
eq.(9) with $\Gamma=2-\frac{2q}{\kappa}$.

{\bf Data availability statement} No data were created or analysed
in this study

{\bf Acknowledgement}

 A part of the results of this paper is
contained in JH's bachelor thesis. JH thanks Dr. Aleksander Kozak
for supervising the thesis. ZH thanks Hytham Alwrekat for
discussions on some  topics of this paper. Both authors thank an anonymous referee for insightful remarks which led to substantial  changes in the manuscript.

\end{document}